\documentstyle[12pt]{article}
\newcommand{\be}{\begin{equation}}
\newcommand{\ee}{\end{equation}}
\newcommand{\ba}{\begin{eqnarray}}
\newcommand{\ea}{\end{eqnarray}}
\newcommand{\baa}{\begin{eqnarray*}}
\newcommand{\eaa}{\end{eqnarray*}}
\newcommand{\bb}{\begin{thebibliography}}
\newcommand{\eb}{\end{thebibliography}}
\newcommand{\ci}[1]{\cite{#1}}
\newcommand{\bi}[1]{\bibitem{#1}}
\newcommand{\lab}[1]{\label{#1}}
\newcommand{\re}[1]{(\ref{#1})}

\newcommand\fac[2]{\mbox{$\frac{#1}{#2}$}}

\newcommand\DGM{XX^{th} }

\begin{document}

\begin{titlepage}
\begin{flushright}
  UdeM-LPN-TH75  \\
  Revised version
\end{flushright}

\vspace*{30mm}

\begin{center}

{\LARGE Exactly Solvable Potentials and Quantum Algebras%
\footnote{Published in Phys.Rev.Lett., vol.69, n.3 (1992) 398-401.}}
\vspace{10mm}

{\large Vyacheslav Spiridonov%
\footnote{On leave from the Institute for Nuclear Research,
Moscow, Russia}$^,$%
\footnote{E-mail address: spiridonov@lps.umontreal.ca}
}

\medskip

{\em Laboratoire de Physique Nucl\' eaire,
Universit\' e de Montr\' eal, \par
C.P. 6128, succ. A, Montr\' eal, Qu\' ebec, H3C 3J7, Canada}

\end{center}

\vspace*{12mm}
\begin{abstract}

A set of exactly solvable one-dimensional quantum mechanical potentials
is described. It is defined by a finite-difference-differential equation
generating in the limiting cases the Rosen-Morse, harmonic, and
P\"oschl-Teller potentials. General solution includes Shabat's
infinite number soliton system and leads to raising and lowering operators
satisfying $q$-deformed harmonic oscillator algebra. In the latter case energy
spectrum is purely exponential and physical states form a reducible
representation of the quantum conformal algebra $su_q(1,1)$.

\medskip

\end{abstract}
\end{titlepage}

\newpage

Lie algebras are among the cornerstones of modern physics. They have
enormous number of applications in quantum mechanics and, in particular,
put an order in classification of exactly solvable potentials.
"Quantized", or $q$-deformed, Lie algebras (also loosely called
quantum groups) are now well established objects in mathematics \ci{r1}.
Their applications were found in two-dimensional integrable models
and systems on lattices. However, despite of much
effort quantum algebras do not yet penetrate into physics on a large scale.
In this paper we add to this field and show that a $q$-deformed
harmonic oscillator algebra \ci{r2} may have straightforward meaning as the
spectrum generating algebra of the specific one-dimensional
potential with exponential spectrum. This result shows that group-theoretical
content of exactly solvable models is not bounded by the standard Lie theory.

Recently Shabat analyzed an infinite chain of reflectionless potentials and
constructed an infinite number soliton system \ci{r3}.
The limiting potential decreased slowly at space infinities and obeyed
peculiar self-similar behavior. We will present corresponding
results in slightly different notations. We denote space variable by $x$
and introduce $N$ superpotentials $W_n(x)$ satisfying the
following set of second order differential equations
\be
(W_n^\prime+W_{n+1}^\prime+W_n^2-W_{n+1}^2)^\prime=0,
\quad n=0,\dots , N-1
\lab{e1}
\ee
where primes denote derivative w.r.t. $x$. Taking first integrals
\be
W_n^\prime+W_{n+1}^\prime+W_n^2-W_{n+1}^2 =k_{n+1},
\lab{e2}
\ee
where $k_n$ are some constants, we define $N+1$ Hamiltonians
\be
2H_n=p^2+U_n(x), \qquad \qquad \qquad p\equiv -id/dx,
\lab{e3}
\ee
$$U_0(x)=W_0^2-W_0^\prime+k_0,\qquad U_{n+1}(x)=U_n(x)+2W_n^\prime (x).$$
An  arbitrary energy shift parameter $k_0$ enters all potentials $U_n(x)$.

Notorious supersymmetric Hamiltonians are obtained by unification of any
two successive pairs $H_n, H_{n+1}$ in a diagonal $2\times 2$ matrix \ci{r3a}.
Analogous construction for
the whole chain \re{e3} was called an order $N$ parasupersymmetric quantum
mechanics \ci{r4,r5}. In the latter case relations \re{e1} naturally arise
as the diagonality conditions of the general $(N+1)\times (N+1)$-dimensional
parasupersymmetric Hamiltonian. We do not use here these algebraic
constructions and consider operators $H_n$ on their own ground.

If $W_n(x)$'s do not have severe singularities
then the spectra of operators \re{e3} may differ only by a finite number
of lowest levels. Under the additional condition that the functions
\be
\psi_{0}^{(n)}(x)=e^{ -\int^x W_n(y)dy}
\lab{e4}
\ee
belong to the Hilbert space ${\cal L}_2$ one obtains first $N$
eigenvalues of the Hamiltonian $H_0$
\be
H_0\,\psi^{(0)}_n(x)=E_n\,\psi^{(0)}_n(x),
\qquad E_n=\fac12\sum_{i=0}^n k_i, \quad n=0,1,\dots ,N-1,
\lab{e5}
\ee
where subscript $n$ numerates levels from below.
In this case \re{e4} represents ground state wave function
of $H_n$ from which one can determine lowest excited states of $H_j, j<n$,
e.g., eigenfunctions of $H_0$ are given by
\be
\psi^{(0)}_n(x)\propto (p+iW_0)(p+iW_1)\dots (p+iW_{n-1})\, \psi_{0}^{(n)}.
\lab{e5a}
\ee
Any exactly solvable discrete spectrum problem can be represented in the
form \re{e2}-\re{e5a}. Sometimes it is easier to solve Schr\"odinger
equation by direct construction of the chain of associated Hamiltonians
\re{e3} -- this is the essence of so called factorization method [7-9].
For the problems with only $N$ bound states there does not exist $W_N(x)$
making $\psi^{(N)}_{0}$ normalizable. For example, if $W_N(x)=0$, then $H_n$
has exactly $N-n$ levels, the potential $U_n(x)$ is reflectionless and
corresponds to $(N-n)$-soliton solution of the KdV-equation.

Let us consider potentials
which support infinite number of bound states, $N=\infty$.
In this case one can derive from \re{e2} the following differential
equations involving only one derivative and a tail of $W_n$'s
\be
W_i^\prime(x)+W^2_i(x)+\sum_{j=1}^\infty (-1)^j (2W^2_{i+j}(x)+k_{i+j})=0.
\lab{e6}
\ee
A question of convergence of the infinite sum is delicate and
requires special consideration in each case. Evident condition
$W_\infty (x)=W^\prime_\infty (x)=0$, which is still
related to the soliton dynamics, is necessary for rigorous justification
of \re{e6}. Here we shall operate with formal series and assume that
initial chain \re{e2} always may be recovered by adding \re{e6} for $i=n$ and
$i=n+1$. In order to find infinite number of superpotentials
$\{ W_i\}$ from \re{e6} one has to relate them to one unknown
function via some simple rule.
Following Ref.\ci{r3} we take the Ansatz
\be
W_i(x)=q^i W(q^i x),
\lab{e8}
\ee
which yields the equation
\be
W^\prime(x)+W^2(x)-\gamma^2 +
2\sum_{j=1}^\infty (-1)^j q^{2j} W^2(q^j x)=0,
\lab {e9}
\ee
where $\gamma^2=-\sum_{j=1}^\infty (-1)^j k_j$.
Note that reality of superpotentials
does not necessarily restrict parameter $q$ to be real -- this
will appear later. From  \re{e9}
it is easy to derive eqs. \re{e6} and \re{e2} with
\be
k_{i+1}=\gamma^2 (1+q^2) q^{2i},\quad i\geq 0.
\lab{e11}
\ee
The following computation
\be
\gamma^2=-\sum_{j=1}^\infty (-1)^j k_j=
\gamma^2(1+q^2)\sum_{j=0}^\infty (-1)^j
q^{2j}\equiv \gamma^2
\lab{e12}
\ee
shows that $\gamma^2$ is completely arbitrary parameter
(an energy scale) and \re{e11} is a
self-consistent definition of the constants $k_i$. Derivation \re{e12}
is valid only at $|q|<1$, which was the restriction of Ref.\ci{r3},
but if \re{e8} and \re{e11} are taken as the basic substitutes for
\re{e2} then by definition $\gamma^2$  is arbitrary and there are no
essential restrictions on $q$ up to now.

Eq. \re{e9} has certain relation to quantum algebras
\ci{r1} and corresponding $q$-analysis \ci{r9}. In order to see this we
first introduce a scaling operator $T_q$ obeying group law
\be
T_q f(x)=f(qx),\quad T_q T_p=T_{qp},\quad T^{-1}_q=T_{q^{-1}},
\quad T_1=1.
\lab{e17}
\ee
Then \re{e9} can be rewritten as
\ba
W^\prime(x)-W^2(x)&=&\gamma^2-2\sum_{j=0}^\infty
(-1)^j (q^2T_q)^j W^2(x)  \nonumber  \\
\medskip
&=&\gamma^2-2(1+q^2 T_q)^{-1}W^2(x).
\lab{e18}
\ea
Multiplying \re{e18} from the l.h.s. by $(1+q^2 T_q)$ we obtain
finite-difference-differential equation defining $W(x)$
\be
W^\prime(x)+W^2(x)+qW^\prime (qx)-q^2 W^2(qx)=\gamma^2 (1+q^2),
\lab{e18a}
\ee
which is nothing else than the first iteration of superpotentials.
The whole infinite chain \re{e2} is thus generated by \re{e18a}.
This observation removes ambiguities arising in \re{e9} due to the
convergence problems.

Let us try to find quantum mechanical spectrum generated by the self-similar
potential $U_0(x)$ associated to \re{e18a}. Suppose that eigenfunctions \re{e4}
are normalizable. Then potential $U_{i+1}(x)$ contains one eigenvalue less
than $U_i(x)$, i.e. there should be the following ordering of levels
\be
E_0<E_1<\dots<E_\infty,\quad
E_n=\fac12 \sum_{i=0}^n k_i=-\fac12 \gamma^2 {1+q^2 \over 1-q^2} q^{2n},
\lab{e13}
\ee
where we chose undefined constant $k_0$ to be
$k_0=-\gamma^2(1+q^2)/(1-q^2)$. At negative $\gamma^2$ it is not possible to
fulfil the ordering and at positive $\gamma^2$ the parameter $q$
should be real and lie in one of the regions $|q|<1$ or $|q|>1$.
Taking the normalization $\gamma^2=\omega^2 |1-q^2|/(1+q^2)$
and denoting $|q|=\exp {(\pm \eta/2)},\; \eta>0,$
we arrive at exponentially small or large bound energy spectrum
\be
E_n=\mp\,\fac12 \, \omega^2 \,e^{\mp \eta n}.
\lab{e14}
\ee

What type of potentials these spectra would correspond to? In order to
know this one should solve equation \re{e18a}. Then everything crucially
depends on the normalizability of $\psi_0^{(0)}$ in \re{e4}
because all other wave functions $\psi^{(n)}_0$ are related to it by scaling.
Normalizability is insured if $W(x)$ is a continuous function positive
at $x\to +\infty$ and negative at $x\to -\infty$. Under such conditions $W(x)$
has at least one zero and we choose corresponding point to be $x=0$, i.e.
$W(0)=0.$ Eq. \re{e18a} now automatically leads to $W(-x)=-W(x)$ and below
we restrict $q$ to be semipositive.
Let us find solution of \re{e18a} in the Taylor series form near the zero.
Substituting an expansion
$W(x)=\sum_{i=1}^\infty c_i x^{2i-1}$ into \re{e9} we obtain
the following recursion relation for the coefficients $c_i$
\be
c_i= {q^{2i}-1 \over q^{2i}+1}\,{1\over 2i-1}\,
\sum_{m=1}^{i-1} c_{i-m} c_m,\quad i\geq 2, \quad c_1=\gamma^2 ,
\lab{e16}
\ee
which at $q=0,\, \gamma=1 $ generates Bernoulli numbers
$B_{2i}$, $c_i=2^{2i}(2^{2i}-1) B_{2i}/(2i)!$.
One may say that \re{e16} defines $q$-analogs of
the Bernoulli numbers $[B_i]_q$. Equation \re{e16} works well for all
values of $q$.
At $q<1$ it describes $q$-deformation of the hyperbolic tangent, since
at $q=0$ one has
\be
W^\prime +W^2=\gamma^2, \qquad W(x)=\gamma \tanh \gamma x ,
\lab{e15}
\ee
which is one-level (soliton) superpotential associated to the
Rosen-Morse problem. At $q>1$ one has $q$-deformation of
the trigonometric tangent which is recovered in the limit $q\to\infty$,
\be
W^\prime -W^2=\gamma^2, \qquad W(x)=\gamma \tan \gamma x .
\lab{e15a}
\ee
This superpotential creates an infinite-level P\"oschl-Teller potential
$U_1(x)$ with the restricted region of coordinate definition:
$-\pi < 2\gamma x < \pi$. On this finite cut $U_0(x)=0$ presents an
infinitely deep potential well. If one sets $\gamma=0$
simultaneously with $q$ or $q^{-1}$ then conformal superpotentials,
$W(x)=\pm 1/x$, are emerging. Finally, at $q=1$ one gets a standard
harmonic oscillator problem.

If $q\ne 0, 1, \infty,$ there is no analytical expression for $W(x)$ but
some general properties of this function may be found along the analysis of
Ref.\ci{r1}, where it was proven that for $q<1$ superpotential is
positive at $x=+\infty$. In this case
required normalizability condition is fulfilled and relation \re{e14} with
upper signs really corresponds to physical spectrum.

At $q>1$ the radius of convergence of the series
defining $W(x)$ is finite, $r_c<\infty$. From inequalities
$${\gamma^2\over\omega^2}\equiv {q^2-1\over q^2+1}\,
< \, {q^{2i}-1\over q^{2i}+1}\,<\, 1, \qquad i>1$$
we have $0<c_i^{(1)}< c_i<c_i^{(2)}$, where $c_i^{(1,2)}$ are defined
by the rule \re{e16} when $q$-factor on the r.h.s. is
replaced by $\gamma^2/\omega^2$ and $1$ respectively
$(c_1^{(1,2)}=c_1)$. As a result, $1<2\gamma r_c/\pi< \omega/\gamma$,
which  means that $W(x)$ is smooth only on some cut at the ends of which
it has singularities. From the basic relation \re{e18a} it follows that
there is an infinite number of simple "primary" and "secondary" poles.
The former ones have residues equal to $-1$ and their location points $x_m$
tend to $\pi (m+1/2)/\gamma,\, m\in Z,$ at $q\to\infty$. "Secondary" poles
are situated at $x=q^n x_m,\, n\in Z^+,$
and corresponding residues are defined by some algebraic equations.
We are thus forced to consider Shr\"odinger operators \re{e3} on a cut
$[-x_1,x_1]$ and impose boundary conditions $\psi_n^{(i)}(\pm x_1)=0$
although the potential $U_0(x)$ is finite at $x=\pm x_1$.
The structure of $W(x)$ leads to $\psi_0^{(0)}(\pm x_1)=0$, i.e. $\psi_0^{(0)}$
is true ground state of $H_0$. Note, however, that the spectrum
$E_n$ for such type of problems can not grow faster than $n^2$ at $n\to \infty$
in apparent contradiction with \re{e14}. This discrepancy is
resolved by observation that already $W_1(x)=qW(qx)$
has singularities on the interval $[-x_1, x_1]$ so that only $H_0$ and
$H_1$ are isospectral in the chain \re{e3}. Hence, the positive signs case
of \re{e14} does not correspond to real physical spectrum of the model.

The number of deformations of a given function is not limited.
The crucial property preserved by the presented above $q$-curling
is the property of exact solvability of "undeformed" Rosen-Morse,
harmonic oscillator, and P\"oschl-Teller potentials.
It is well known that potentials at infinitely small and exact
zero values of a parameter may obey completely different spectra.
In our case, deformation with $q<1$ converts one-level problem
\re{e15} with $E_0=-\gamma^2/2$ into the infinite-level one with
exponentially small energy eigenvalues \re{e14}.
Whether one gets exactly solvable potential
at $q>1$ is an open question but this is quite plausible because at $q=\infty$
a problem with known spectrum $E_n=\gamma^2 (n+1)^2/2$ arises.

In standard dynamical symmetry approach Hamiltonian of a system is
supposed to be proportional either to Casimir operator or to polynomial
combination of the generators of some Lie algebra \ci{r7,r10}. As a result,
energy eigenvalues are determined by rational functions of
quantum numbers. This means that one does not go out of the
universal enveloping algebra. $q$-Deformation of the
universal algebra works with functions (exponentials) of generators
and, as it was announced, accounts for the presented exponential spectra.

Indeed, substituting superpotentials \re{e8} into relation \re{e5a}
one finds raising and lowering operators
$$\psi_{n\pm 1}^{(0)}\propto A^{\pm} \psi_n^{(0)}, \qquad\qquad
H_0=\fac12 (A^+ A^- -{1+q^2\over 1-q^2}\gamma^2),$$
\be
A^+=q^{1/2} (p+iW(x))T_q, \qquad
A^-=q^{-1/2} T_q^{-1}(p-iW(x)),
\lab{e19}
\ee
For real $q$ and $\gamma$ the operators $A^{\pm}$ are hermitian conjugates of
each other. Eq. \re{e18a} insures the following $q$-commutation relations
\be
A^- A^+ - q^2 A^+ A^- =\gamma^2 (1+q^2),\qquad
H_0 A^{\pm}=q^{\pm 2} A^{\pm} H_0.
\lab{e20}
\ee
Introduction of formal number operator
\be
N={\ln H_0/ E_0\over \ln q^2}, \qquad
N\, \psi_n^{(0)} =n\, \psi_n^{(0)}, \qquad
[N,A^{\pm}]=\pm A^{\pm}
\lab{e21}
\ee
completes the definition of $q$-deformed harmonic oscillator algebra
in the particular form \ci{r2}.
The quantum conformal algebra $su_q(1,1)$ is realized as follows \ci{r10a},
$$K^+=({q\over \gamma(1+q^2)} q^{-N/2} A^+)^2,\qquad
K^-=(K^+)^{\dag},
\qquad K_0=\fac12 (N+\fac12), $$
\be
[K_0, K^{\pm}]=\pm K^{\pm}, \qquad [K^+,K^-]=-
{q^{4K_0}-q^{-4K_0}\over q^2-q^{-2}},
\lab{e22a}
\ee
i.e. it is a dynamical symmetry algebra of the model.
Generators $K^{\pm}$ are parity invariant and therefore even and odd wave
functions belong to different irreducible representations of $su_q(1,1)$.

In order to generalize basic equation \re{e18a} we introduce an additional
parameter $s$ into the superpotential, $W=W(x,s)$, and assume that
$T_q$ in \re{e19} is a generalized shift operator
\be
T_q W(x,s)=W(qx+a,s+1),\qquad 
\lab{e24}
\ee
where $q$ and $a$ are parameters of affine transformation.
Although $A^+$ is not hermitian conjugate of $A^-$ any more, we
force them to obey $q$-oscillator type algebra
$$A^- A^+ - q^2 A^+ A^- = C(s),\qquad A^{\pm} C(s)=C(s\pm 1) A^{\pm},$$
where $C$ is some function of $s$. Resulting equation
for the superpotential
\be
W^\prime(x,s-1)+qW^\prime(qx+a,s) +W^2(x,s-1) -q^2W^2(qx+a,s)=C(s),
\lab{e26}
\ee
may be called the generalized shape-invariance condition (cf. \ci{r8}).


We define a Hamiltonian $H$ as follows
\be
H=\fac12 A^+ A^- + F(s),\qquad  q^2 F(s)-F(s-1)=\fac12 C(s),
\lab{e26a}
\ee
where finite-difference equation for the function $F(s)$ is found from
the braiding relations $HA^{\pm}=q^{\pm 2}A^{\pm}H$.
Now it is easy to generalize formula \re{e13}.
Suppose that a wave function $\psi_0$, $A^-\psi_0=0,$ is
normalizable. Then a tower of higher states $\psi_n\propto (A^+)^n \psi_0$
gives energy spectrum
\be
E_n=F(s)+\fac12 \sum_{i=1}^n q^{2(i-1)} C(s+i)= q^{2n} F(s+n),
\lab{e27}
\ee
which can be found by purely algebraic means.
If for some $n=N$ normalizability of $\psi_n$ is broken then $H$ has only
$N$ discrete levels. In the above presentation we
chose the simplest form of $s$-parameter transformation under the action
of $T_q$-operator. One can easily generalize formula \re{e27} for arbitrary
change of variable $s$ in \re{e24}, $s\to f(s)$.

To conclude, in this paper we have described an exactly solvable quantum
mechanical problem where quantum algebra $su_q(1,1)$ acts on the discrete
set of energy eigenstates scaling their eigenvalues by the constant factor.
In the original version of differential
geometric applications of quantum Lie algebras an underlying space
was taken to be non-commutative ("quantum plane") and deformation parameter $q$
was measuring deviations from normal analysis (see, e.g., Ref.\ci{r11}).
Here we have commutative space and standard one-dimensional
quantum mechanics but the potential is very peculiar.
It represents $q$-deformation
of exactly solvable potentials so that the spectrum remains to be
known but it acquires essentially functional character.

It is interesting to know the most general exactly solvable $q$-deformed
potential. One of the approaches to this problem consists
in repetition of the trick described in Ref.\ci{r12}. Namely, one can
take as particle's wave function a $q$-hypergeometric function multiplied
by some weight factor. This would correspond
to the transformation of $q$-hypergeometric equation to the form
of standard Schr\"odinger equation for some potential.
Another path to $q$-deformation of known models is given by the eq. \re{e26}
which may have solutions generalizing those found by the old
factorization technique at $q=1,\, a=0$.

Two final remarks are in order. First, affine transformations appearing
in \re{e26} may be used for the definition of $q$-deformed
supersymmetric quantum mechanics \ci{r13}. Second, at complex values of $q$
one has meaningful dynamical systems which are exactly solvable when
$q$ is a root of unity \ci{r14}.

The author is indebted to A.Shabat for acquainting with his
paper prior to publication and for relevant remarks.
This work was supported by the NSERC of Canada.

\newpage

\bb{22}
\bi{r1} V.G.Drinfeld, Sov.Math.Dokl. {\bf 32}, 254 (1985);
        M.Jimbo, Lett.Math.Phys. {\bf 10}, 63 (1985); {\bf 11},
        247 (1986); N.Yu.Reshetikhin, L.A.Takhtajan, and
     L.D.Faddeev, Algebra i Analiz, {\bf 1}, 178 (1989).

\bi{r2} L.C.Biedenharn, J.Phys. {\bf A22}, L873 (1989);
        A.J.Macfarlane, J.Phys. {\bf A22}, 4581 (1989).

\bi{r3} A.Shabat, Inverse Problems, {\bf 8}, 303 (1992).

\bi{r3a} E.Witten, Nucl.Phys. {\bf B188}, 513 (1981).

\bi{r4} V.A.Rubakov and V.P.Spiridonov,
Mod. Phys. Lett. {\bf A3}, 1337 (1988);
V.Spiridonov, {\it in}: Proc. of the $\DGM$ DGM Conf.,
New York, USA, 3-7 June 1991. Eds. S.Catto and A.Rocha (World
Scientific, 1992) p. 622.

\bi{r5}  S.Durand, M.Mayrand, V.P.Spiridonov, and L.Vinet, Mod.Phys.Lett.
{\bf A6}, 3163 (1991).

\bi{r6} L.Infeld and T.E.Hull, Rev.Mod.Phys. {\bf 23}, 21 (1951).

\bi{r7} W.Miller, Jr., Lie Theory and Special Functions (Academic Press,
        1968).

\bi{r8} L.E.Gendenstein, Pis'ma ZhETF {\bf 38}, 299 (1983).

\bi{r9} For a review and list of references see R.Floreanini and
        L.Vinet, {\it in}: Proc. of the ${\it II^{nd}}$ Int. Wigner
        Symp., Goslar, Germany, 16-20 July 1991, to be published.


\bi{r10} Y.Alhassid, F.G\"ursey, and F.Iachello,
        Ann.Phys. (N.Y.) {\bf 148}, 346 (1983);    \\
        M.Moshinsky, C.Quesne, and G.Loyola, Ann.Phys. (N.Y.) {\bf 198},
        103 (1990).

\bi{r10a} P.P.Kulish and E.V.Damaskinsky, J.Phys. {\bf A23}, L415 (1990).

\bi{r11} J.Wess and B.Zumino, Nucl.Phys. (Proc.Suppl.) {\bf B18},
         302 (1990); \\
         B.Zumino, Mod.Phys.Lett. {\bf A6}, 1225 (1991).

\bi{r12} G.A.Natanzon, Vestnik Leningrad Univ. {\bf 10}, 22 (1971);   \\
         O.V.Bychuk and V.P.Spiridonov, Mod.Phys.Lett.
         {\bf A5}, 1007 (1990).

\bi{r13} V.Spiridonov, Mod.Phys.Lett. {\bf A7}, 1241 (1992).

\bi{r14} A.Shabat and V.Spiridonov, unpublished.

\eb
\end{document}